\documentclass[
reggeiaps,prd,
nofootinbib,
superscriptaddress,
showpacs,ligh
tightenlines,
]{revtex4}

\usepackage{amsmath}
\usepackage{amssymb}
\usepackage{bm}
\usepackage{color,graphicx}

\begin{document}

\title{On the Observability of the Quark Orbital Angular Momentum Distribution}

\author{Aurore Courtoy} 
\email{aurore.courtoy@ulg.be}
\affiliation{IFPA, AGO Department, Universit\'e de Li\`ege, B\^at. B5, Sart Tilman B-4000 Li\`ege, Belgium
\\ and Laboratori Nazionali di Frascati, INFN, Frascati, Italy.}

\author{Gary R.~Goldstein} 
\email{gary.goldstein@tufts.edu}
\affiliation{Department of Physics and Astronomy, Tufts University, Medford, MA 02155 USA.}

\author{J. Osvaldo Gonzalez Hernandez}
\email{jog4m@virginia.edu}
\affiliation{Istituto Nazionale di Fisica Nucleare (INFN) - Sezione di Torino
via P. Giuria, 1, 10125 Torino, ITALY, Italy}

\author{Simonetta Liuti }
\email{sl4y@virginia.edu}
\affiliation{University of Virginia - Physics Department,
382 McCormick Rd., Charlottesville, Virginia 22904 - USA \\ and Laboratori Nazionali di Frascati, INFN, Frascati, Italy.} 

\author{Abha Rajan}
\email{ar5xc@virginia.edu}
\affiliation{University of Virginia - Physics Department,
382 McCormick Rd., Charlottesville, Virginia 22904 - USA}


\begin{abstract}
We argue that due to Parity constraints, the helicity combination of the purely momentum space counterparts of the Wigner distributions -- the generalized transverse momentum distributions --
that describes the configuration of an unpolarized quark in a longitudinally polarized nucleon, 
can enter the deeply virtual Compton scattering amplitude only through matrix elements involving a final state interaction.
The relevant matrix elements in turn involve light cone operators projections in the transverse direction, or they appear 
in the deeply virtual Compton scattering amplitude at twist three.
Orbital angular momentum or the spin structure of the nucleon was a major reason for these various distributions and amplitudes to have been introduced. 
We show that the twist three contributions associated  to orbital angular momentum
 are related to the target-spin asymmetry in deeply virtual Compton scattering, already measured at HERMES.
\end{abstract}

\maketitle

\begin{figure}
\hspace{-0.5cm}
\includegraphics[width=8cm]{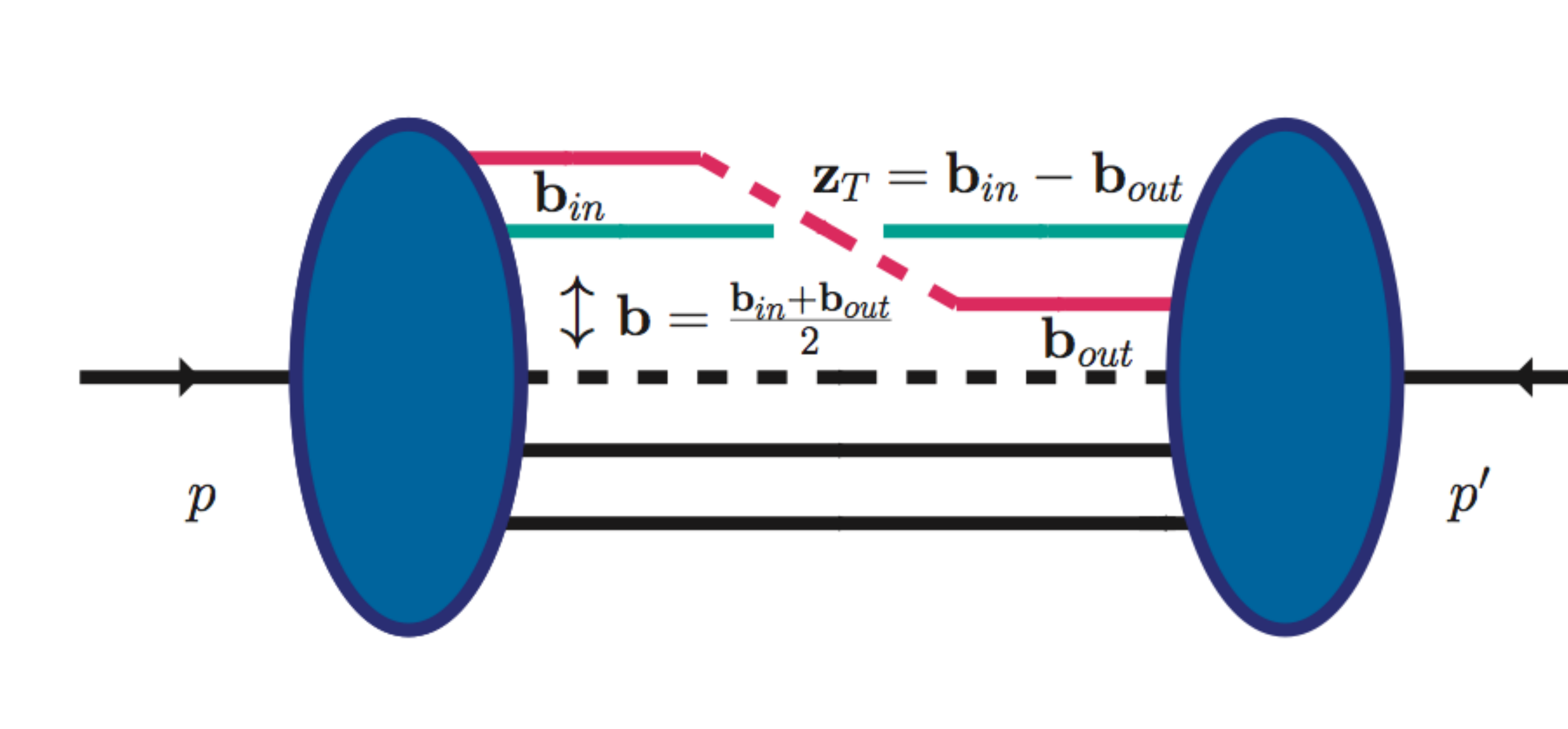}
\hspace{2cm}
\includegraphics[width=8.cm]{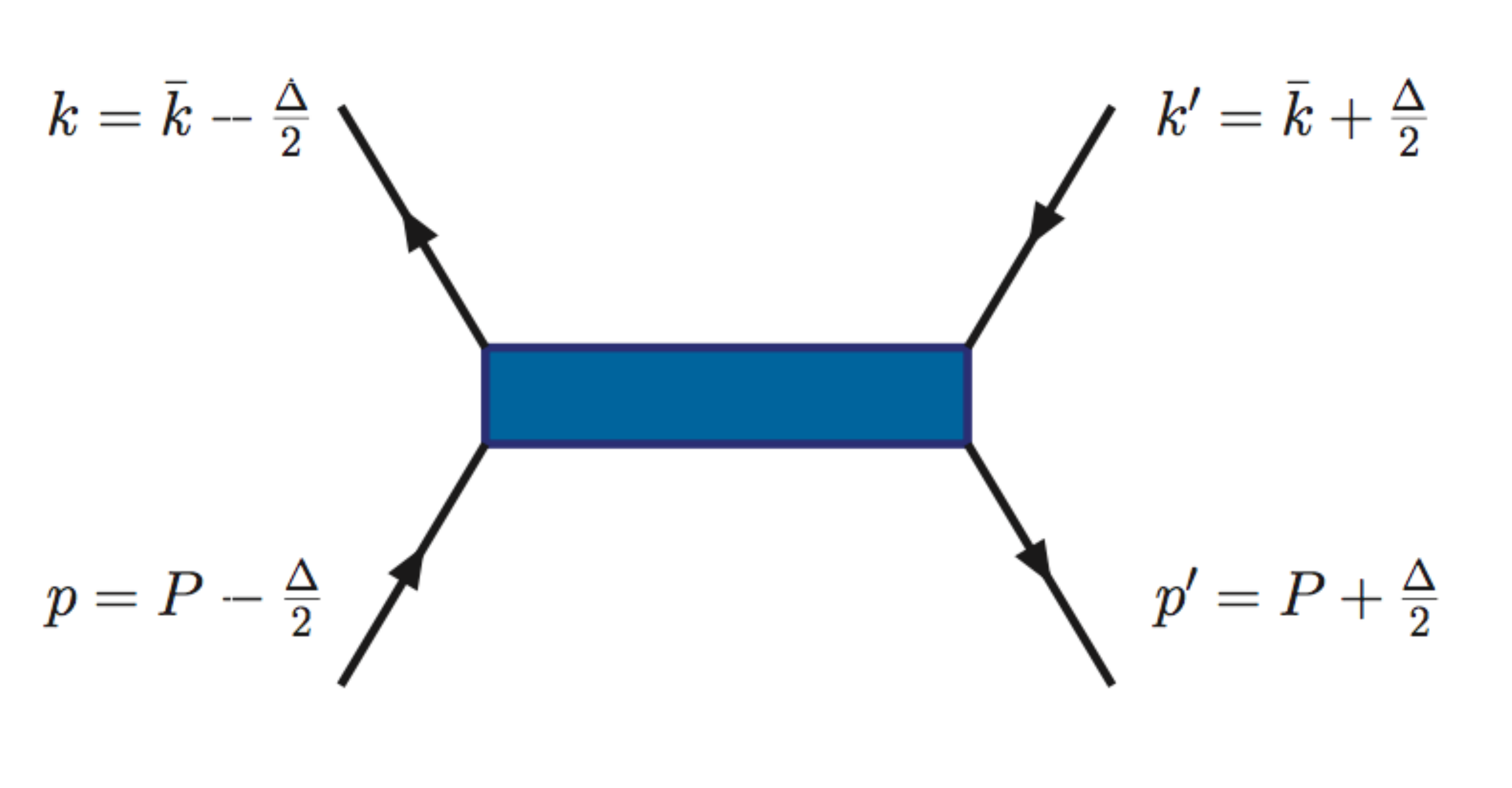}
\caption{{\bf (a)} Left: Correlation function for a GTMD; {\bf (b)} quark-proton scattering in the $u$-channel.}
\label{fig1_f14}
\end{figure}

%
\noindent {\bf 1.} Considerable attention has been devoted to the partons' Transverse Momentum Distributions (TMDs), to the Generalized Parton Distributions (GPDs), and to finding a connection between the two \cite{Bur,Metz1,Metz2}. 
TMDs are distributions of different spin configurations of quarks and gluons within the nucleon whose longitudinal and transverse momenta can be accessed in Semi-Inclusive Deep Inelastic Scattering (SIDIS). 
GPDs are real amplitudes for quarks or gluons being probed in a hard process and then returning to reconstitute a scattered  nucleon. 
They are accessed through exclusive electroproduction of vector bosons along with the nucleon. In each case there is a nucleon matrix element of bilinear, non-local quark or gluon field operators. In principle both TMDs and GPDs are different limits of Wigner distributions, {\it i.e.} the  phase space distributions in momenta and impact parameters. The purely momentum space form of those are the Generalized TMDs (GTMDs). 
GTMDs correlate hadronic states with same parton longitudinal momentum, $x$ (assuming zero skewness), different relative transverse distance, ${\bf z}_T={\bf b}_{in}-{\bf b}_{out}$, between the struck parton's initial and final (out) states, and same average transverse distance, ${\bf b}=({\bf b}_{in}+{\bf b}_{out})/2$, of the struck parton with respect to the center of momentum \cite{Soper77} (Figure \ref{fig1_f14}a).

Understanding the angular momentum or spin structure of the nucleon is a major reason for these various distributions and amplitudes to have been introduced. 
%
Recently, specific GTMDs and Wigner distributions were studied that are thought to
be related to the more elusive component of the  angular momentum sum rule, which is partonic Orbital Angular Momentum (OAM) \cite{Vento,Brodsky,LorPas,Yuan}. Such theoretical efforts have been developing in parallel with  the realization that the leading twist contribution to the angular momentum sum rule comes from transverse spin \cite{Ji_Xiong}, while longitudinal angular momentum, and consequently orbital angular momentum, can be associated with twist three partonic components.
The GTMD that was proposed to describe OAM appears in the parametrization of the (leading order) vector, $\gamma^+$, component of the unintegrated quark-quark correlator for the proton given in \cite{Metz2} as,
\begin{equation}
\label{S_L}
 \bar{u}(p',\Lambda') \frac{i \sigma^{ij} \bar{k}_T^i \Delta_T^j}{M^2} u(p,\Lambda) F_{14}  \propto \langle {\bf S}_L \cdot \bar{\bf k}_T \times {\bf \Delta}_T \rangle.
 \end{equation}
where $(p, \Lambda), (p',\Lambda')$ are the proton's initial and final momentum and helicity, $k_T$ and $\Delta_T$ are the quarks' average and relative momenta, respectively, and $F_{14}$ is the GTMD defined according to the classification scheme of \cite{Metz2}.  Eq.(\ref{S_L}) appears to represent the distribution of an unpolarized quark in a longitudinally polarized proton ($\rho_{LU}$ in Ref.\cite{LorPas}).  
However, we notice that its observability is inconsistent with several physical properties, namely: 

\vspace{0.3cm}
\noindent  {\it  i)} it drops out of the formulation of both GPDs and TMDs, so that it cannot be measured;

\noindent {\it ii)} it is Parity-odd. Notice, in fact, how
this structure is at variance with the more familiar $\langle {\bf S}_T \times \bar{\bf k}_T \rangle$ term, defining the TMD Sivers function \cite{Sivers}, which is a Parity even observable with an analogous asymmetry but for transverse spin. It also differs from the definition of OAM based on the TMD $h_{1T}^\perp$ \cite{Avakian,Ma}, in that the latter also exhibits a transverse spin structure.   

%

\noindent {\it iii)} it is non zero only for imaginary values of the quark-proton helicity amplitudes.      
\vspace{0.3cm}

In order to develop a more concrete understanding of OAM that could lead to the definition of specific observables, it is important to 
examine the newly proposed partonic configurations and their connection with the quark-proton helicity or transverse spin amplitudes. 
In this paper, after showing the constraints on the recently studied GTMDs from the invariance under Parity transformations, 
we perform a thorough analysis of their helicity/transverse spin structure. We suggest that the helicity structure giving rise to OAM 
 is described by twist  three distributions only. 
Our findings corroborate the interpretation of the spin sum rule including OAM originally given in \cite{Polyakov} (see also \cite{Penttinen,Hatta}). 
Using the formalism proposed in Ref.~\cite{Metz2,BelMuel}, we identify  the integrals over transverse momentum of the twist three GTMDs with twist three GPDs. The latter enter  the $\sin 2\phi $ modulation of the target-spin asymmetry, $A_{UL}$, in DVCS which has already been measured at both HERMES~\cite{HERMES} and JLab~\cite{pisano_privcomm}.  We present here predictions for $A_{UL}^{\sin 2  \phi}$ using the GPD set of Ref.~\cite{GGL,GGL_FF}.
This is the first proposal  for a direct experimental access to orbital angular momentum.

\vspace{0.3cm}
\noindent {\bf 2.} 
%
In Ref.~\cite{Metz2} it was found that with Parity invariance, time reversal invariance, and Hermiticity there are 16 independent complex GTMDs for the quark-nucleon system, corresponding to 16 helicity amplitudes for quark-nucleon elastic scattering. 
However, we know that for elastic 2-body scattering of two spin $1/2$ particles there will only be 8 independent amplitudes. This follows from implementing Parity transformations on the helicity amplitudes in the 2-body Center of Mass (CM)  frame \cite{JacobWick} where all the incoming and outgoing particles are  confined to a plane. In this plane the Parity transformation flips all momenta but it does not change the relation among the momentum components. In any other frame there will still be 8 independent amplitudes, although they may be in linear combinations with kinematic factors that {\em appear} to yield 16. 
The counting of helicity amplitudes in polarization dependent high energy scattering processes was addressed {\it e.g.} Ref.~\cite{FGM}. In order to explain this point, and to investigate its consequences on the  off-forward matrix elements of QCD correlators, we first start by reviewing the helicity structure of the GTMDs from Ref.\cite{Metz2}.

To describe quark-proton scattering as a $u$-channel  two-body scattering process (Fig.\ref{fig1_f14}b)
\[ q^\prime(k^\prime) + N(p) \rightarrow q(k) + N^\prime(p^\prime),  \] 
we choose a Light-Cone (LC) frame, where the {\em average} and {\em relative} 4-momenta $P=(p+p')/2$, $\bar{k}= (k+k')/2$, $\Delta = p'-p= k'-k$, respectively have components specified by,
\begin{subequations}
\label{coord}
\begin{eqnarray}
P & \equiv & \left(P^+, \frac{{\bf \Delta}_T^2 + 4 M^2}{8P^+}, 0 \right)  \\
\bar{k} & \equiv &  \left( xP^+,  k^-, \bar{\bf k}_T  \right)  \\
\Delta & \equiv &  \left( 0 , 0, {\bf \Delta}_T \right)  
\end{eqnarray}
\end{subequations}
where $v \equiv (v^+,v^-, {\bf v}_T)$,  $v^\pm=1/\sqrt{2}(v_o \pm v_3)$, and
for simplicity we have taken the skewness variable, $\xi=0$ since this will not enter our discussion.

The connection between the unintegrated matrix elements defining the GTMDs and the quark-proton helicity amplitudes  
is obtained by considering the quark-proton helicity amplitude (Fig.\ref{fig1_f14} and Ref.\cite{Diehl_rev}),
\begin{eqnarray}
A_{\Lambda' \lambda', \Lambda \lambda} = \int \frac{d z^- \, d^2{\bf z}_T}{(2 \pi)^3} e^{ixP^+ z^- - i \bar{\bf k}_T\cdot {\bf z}_T} \left. \langle p', \Lambda' \mid {\cal O}_{\lambda' \lambda}(z) \mid p, \Lambda \rangle \right|_{z^+=0}, \nonumber \\
\end{eqnarray} 
%
where in the chiral even sector, 
\begin{eqnarray}
\label{tw2_operator}
{\cal O}_{\pm \pm}(z) & = & \bar{\psi}\left(-\frac{z}{2}\right) \gamma^+(1 \pm \gamma_5)  \psi\left(\frac{z}{2}\right). 
\end{eqnarray}
Following the definitions in Ref.~\cite{Metz2}, and making use of the Gordon decomposition, 
\begin{equation}
\label{Gordon}
\overline{U}(p',\Lambda')  \gamma^\mu U(p,\Lambda) = \overline{U}(p',\Lambda') \frac{P^\mu}{M} U(p,\Lambda) + \overline{U}(p',\Lambda')  \frac{i \sigma^{i \, \mu}\Delta_i}{2M} U(p,\Lambda),
\end{equation}
we obtain for the vector case,
\begin{eqnarray}
\label{vector}
W^{\gamma^+}_{\Lambda \Lambda'} & = & 
\frac{1}{2 P^+}  \left[ \overline{U}(p',\Lambda') \gamma^+ U(p,\Lambda) F_{11} 
+ \overline{U}(p',\Lambda') \frac{i \sigma^{i+} \Delta_T^i}{2M} U(p,\Lambda) (2 F_{13} - F_{11})  \right. \nonumber \\
& + & \left.  \overline{U}(p',\Lambda') \frac{i \sigma^{i+} \bar{k}_T^i}{2M} U(p,\Lambda) (2 F_{12} ) +
  \overline{U}(p',\Lambda') \frac{i \sigma^{ij} \bar{k}_T^i \Delta_T^j}{M^2}  U(p,\Lambda) \, F_{14}  \right]  \nonumber \\
& = & \delta_{\Lambda, \Lambda'} F_{11} +  \delta_{\Lambda, -\Lambda'}  \frac{-\Lambda \Delta_1 - i \Delta_2}{2M}  (2 F_{13} - F_{11}) +  
 \delta_{\Lambda,- \Lambda'}  \frac{-\Lambda \bar{k}_1 - i \bar{k}_2}{2M}  (2 F_{12}) + \delta_{\Lambda, \Lambda'} i \Lambda \frac{\bar{k}_1\Delta_2 - \bar{k}_2 \Delta_1}{M^2} F_{14}, 
\end{eqnarray}
and for the axial-vector case,
\begin{eqnarray}
\label{axial}
W^{\gamma^+\gamma_5}_{\Lambda \Lambda'} & = & \frac{1}{2\,{P}^+} 
 \left[ -\overline{U}(p',\Lambda') \frac{i \epsilon_T^{ij} \,\bar{k}_T^i\Delta_T^j}{M^2} \,\frac{2P^+}{2M}  U(p,\Lambda) G_{11} 
+ \overline{U}(p',\Lambda') \frac{i \sigma^{i+}\gamma_5 \Delta_T^i}{2M} U(p,\Lambda) (2 G_{13} )  \right. \nonumber \\
& + & \left.  \overline{U}(p',\Lambda') \frac{i \sigma^{i+} \gamma_5\, \bar{k}_T^i}{2M} U(p,\Lambda) (2 G_{12} ) 
+ \overline{U}(p',\Lambda') i \sigma^{+-} \gamma_5  U(p,\Lambda) \, G_{14}  \right]  \nonumber \\
& = & \delta_{\Lambda, \Lambda'} \Lambda G_{14} +  \delta_{\Lambda, -\Lambda'}  \frac{\Delta_1 + i \Lambda \Delta_2}{2M}  \, 2 G_{13} +  
 \delta_{\Lambda,- \Lambda'}  \frac{\bar{k}_1 + i \Lambda \bar{k}_2}{2M}  \, 2 G_{12} -  \delta_{\Lambda, \Lambda'} i \frac{\bar{k}_1\Delta_2 - \bar{k}_2 \Delta_1}{M^2}  G_{11}, 
 \nonumber \\
\end{eqnarray}

By summing and subtracting the two equations, one obtains the following expressions for the quark-proton helicity amplitudes,
\begin{subequations}
\label{ALambda}
\begin{eqnarray}
A_{\Lambda' +,\Lambda,+} & = &
 \delta_{\Lambda, \Lambda'} (F_{11} + \Lambda G_{14} +   i \Lambda \frac{\bar{k}_1\Delta_2 - \bar{k}_2 \Delta_1}{M^2} F_{14} -  i  \frac{\bar{k}_1\Delta_2 - \bar{k}_2 \Delta_1}{M^2}   G_{11}) + \nonumber \\
& & \delta_{\Lambda, -\Lambda'}  \left[\frac{-\Lambda \Delta_1 - i \Delta_2}{2M}  (2 F_{13} - F_{11}) +  \frac{\Delta_1 - i \Lambda \Delta_2}{2M}  2 G_{13} 
+ \frac{-\Lambda \bar{k}_1 - i \bar{k}_2}{2M}  2F_{12}  + \frac{\bar{k}_1 - i \Lambda \bar{k}_2}{2M} 2G_{12}\right]   \nonumber \\
&& \\
A_{\Lambda' -,\Lambda,-} & = & \delta_{\Lambda, \Lambda'} (F_{11} - \Lambda G_{14} +  i \Lambda \frac{\bar{k}_1\Delta_2 - \bar{k}_2 \Delta_1}{M^2} F_{14} + i  \,  \frac{\bar{k}_1\Delta_2 - \bar{k}_2 \Delta_1}{M^2}  G_{11}) + \nonumber \\ 
& & \delta_{\Lambda', -\Lambda}  \left[\frac{-\Lambda \Delta_1 - i \Delta_2}{2M}  (2 F_{13} - F_{11}) -  \frac{\Delta_1 - i \Lambda \Delta_2}{2M}  2 G_{13} + \frac{-\Lambda \bar{k}_1 - i \bar{k}_2}{2M}  2F_{12}  - \frac{\bar{k}_1 - i \Lambda \bar{k}_2}{2M} 2G_{12}\right], \nonumber \\
\end{eqnarray}
\end{subequations}
We now examine the new contributions, $F_{14}$, and $G_{11}$,
\begin{subequations}
\begin{eqnarray}
\label{F14}
 i  \frac{\bar{k}_1\Delta_2 - \bar{k}_2 \Delta_1}{M^2} F_{14} & = & A_{++,++} + A_{+-,+-} - A_{-+,-+} - A_{--,--} \\
 \label{G11}
 -i  \frac{\bar{k}_1\Delta_2 - \bar{k}_2 \Delta_1}{M^2}  G_{11} & = & A_{++,++} -A_{+-,+-} + A_{-+,-+} - A_{--,--}   
\end{eqnarray}
\end{subequations}
$F_{14}$ describes an unpolarized quark in a longitudinally polarized proton, while $G_{11}$ describes a longitudinally polarized quark in an unpolarized proton.


We reiterate, however, that Parity, imposes limits on the possible polarization asymmetries that can be observed in two body scattering:    
because of 4-momentum conservation and on-shell conditions, $k^2=m^2$, $p^2=M^2$, there are eight variables. Four of those  describe the energy and 3-momentum of the CM relative to a fixed coordinate system, while the remaining four give the energy  and the 3-vector orientation and magnitude of the scattering plane in the CM.
In the CM frame or, equivalently in  the ``lab" frame with the $p$ direction chosen as the $z$-direction, 
the net longitudinal polarization 
 defined in Eq.(\ref{S_L}), is clearly a Parity violating term (pseudoscalar) under space inversion. 
This implies that a measurement of  {\em single longitudinal polarization asymmetries} would violate Parity conservation in an ordinary two body scattering process corresponding to tree level, twist two amplitudes. 
Releasing the partons' on-shell condition implies introducing higher twists in the description of the process \cite{EFP}
We can therefore anticipate that similarly to the TMDs $g^\perp, f_L^\perp, \ldots $ in SIDIS, single longitudinal polarization asymmetries are higher twist objects. 

On the other hand, notice that polarization along the normal to the scattering plane is Parity conserving (scalar) under spatial 
inversion, thus giving rise to SSAs at leading twist \cite{JacRobRos}. 

To be explicit regarding parity constraints consider again the helicity amplitudes for the 2-body process, 
\begin{equation}
A_{\Lambda^\prime,\, \lambda^\prime; \, \Lambda, \, \lambda} : q^\prime(k^\prime,\lambda^\prime) + N(p,\Lambda) \rightarrow q(k,\lambda) + N^\prime(p^\prime,\Lambda^\prime).
\label{helamps}
\end{equation}
Such amplitudes can be written in any Lorentz frame, but in the Center of Momentum frame the  parity relations are simple,
\begin{equation}
A_{-\Lambda^\prime,\, -\lambda^\prime; \, -\Lambda, \, -\lambda}=(-1)^\eta \, A_{\Lambda^\prime,\, \lambda^\prime; \, \Lambda, \, \lambda}^*,
\label{helparity}
\end{equation}
where $\eta =  \Lambda^\prime - \lambda^\prime - \Lambda + \lambda$, the net helicity change. Hence of 16 possible helicity amplitudes, 8 are independent. For chiral even, non-flip nucleon amplitudes there are 2 independent.  These determinations are made in the CoM frame. By Lorentz covariance and 4-momentum conservation, the number of independent amplitudes cannot change. In other frames, {\it e.g.} the light cone frame or the target rest frame, there may appear to be more, yet the extra amplitudes must be linear combinations of the independent ones. In particular
\begin{equation}
A_{\widetilde{\Lambda}^\prime,\, \widetilde{\lambda}^\prime; \, \widetilde{\Lambda}, \, \widetilde{\lambda}}= \sum_{\Lambda^\prime, \cdots} D_{\widetilde{\Lambda}^\prime,\,\Lambda^\prime}^{1/2}(\Omega_N^\prime)  \cdots A_{\Lambda^\prime,\, \lambda^\prime; \, \Lambda, \, \lambda},
\label{heltrans}
\end{equation}
where the $D$ functions are the rotation matrices for the Wigner rotations, $\Omega_N^\prime$, etc.

We see that for $F_{14}$ in Eq.~\ref{F14} and $G_{11}$ in Eq.~\ref{G11} to be non-zero there must be an imaginary part to either $A_{++;++}$ or $A_{+-;+-}$. This will not be the case in the CoM frame, wherein the momenta are coplanar. In order to have a non-vanishing helicity amplitude combination there must be another independent direction. That is provided by twist three amplitudes and corresponding GTMDs, as we show below.

Note also that that in the parametrization of the generalized correlation function of Ref.\cite{Metz2},
\begin{eqnarray}
W_{\Lambda^\prime,\, \Lambda}^{\gamma^+} &=&  {\overline U}(p^\prime, \Lambda^\prime )\left[ \overbrace{\frac{P^+}{M} (A_1^F+xA_2^F -2\xi A_3^F)}^{type 1} +\overbrace{\frac{i\sigma^{+k}}{M} A_5^F + \frac{i\sigma^{+\Delta}}{M} A_6^F}^{type 2} +\overbrace{\frac{P^+i\sigma^{k\,\Delta}}{M^3}(A_8^F+xA_9^F)}^{type 3} \right. \nonumber \\
& & \left. +  \underbrace{\frac{P^+i\sigma^{k\,N}}{M^3}(A_{11}^F+xA_{12}^F) +\frac{P^+i\sigma^{\Delta\,N}}{M^3}(A_{14}^F -2\xi A_{15}^F)}_{type 4} \right] U(p,\Lambda) \nonumber \\
& = & A_{\Lambda' \,+ ; \Lambda \, +}^{[\gamma^+]} + A_{\Lambda' \,- ; \Lambda \, -}^{[\gamma^+]}
\end{eqnarray}
the Dirac structure reduces to the four distinct groups selected above, of which type 1,2 and 4 are Parity even, while type 3, which composes $F_{14}$, is Parity odd 
\cite{SL_talkQCDJlab}
(details of the calculation will be given in \cite{inprogress}). 
Because of the Parity constraints \cite{Metz2}  $F_{14}$ can therefore be non zero only if its corresponding helicity amplitudes combination is imaginary. Hence it cannot have a straightforward partonic interpretation. Integrating over $k_T$ gives zero for $F_{14}$ meaning that this term decouples from partonic angular momentum sum rules. 
The vector GPDs can, in fact, be written in terms of GTMDs as \cite{Metz2},
\begin{eqnarray}
H & = & \int d^2 \bar{\bf k}_T  F_{11} \\
E & = & \int  d^2 \bar{\bf k}_T  \left[ - F_{11} + 2\left( \frac{\bar{\bf k}_T \cdot {\bf \Delta}}{\Delta^2} F_{12} + F_{13} \right) \right]. 
\end{eqnarray}

We conclude that type 3 should not be included in the leading twist parametrization in Eq.(\ref{vector}). A similar argument is valid for the axial vector component (Eq.(\ref{axial})). 

\vspace{0.3cm}
\noindent {\bf 3.} 
In the presence of final state interactions Parity relations apply differently. We will show that  the combination
%
\[ A_{++,++} + A_{+-,+-}  - A_{-+,-+}- A_{--,--}, \]
gets replaced by a similar helicity structure where now the longitudinal spin is crossed into ${\bf \Delta}$, namely  $\langle {\bf S}_L \times {\bf \Delta}_T \rangle$. Notice that this produces a transverse angular momentum component rather than the longitudinal component appearing in Eq.(\ref{S_L}).

The chiral-even twist three components were also parametrized in Ref.\cite{Metz2},
\begin{eqnarray}
\label{tw3_metz}
W^{\gamma^i}_{\Lambda' \Lambda} & = & \frac{1}{2 P^+} \overline{U}(p',\Lambda')  \left[%
\frac{\bar{k}_T^i}{M}  F_{21}  + 
\frac{\Delta_T^i}{M} (F_{22}- 2  F_{26})
 +   \frac{ i\sigma^{ji}\bar{k}_T^j}{M}  F_{27}  +  \gamma^i (2F_{28}) \right. \nonumber \\
 & + &  \left. \frac{M i\sigma^{i+}}{P^+}  F_{23} +   \frac{\bar{k}_T^i}{M} \frac{ i\sigma^{k+}\bar{k}_T^k}{P^+}  F_{24} + \frac{\Delta_T^i}{M} \frac{ i\sigma^{k+}\bar{k}_T^k}{P^+} F_{25}
+     \frac{\Delta_T^i}{P^+} \gamma^+ (2 F_{26})   \right]    U(p,\Lambda), 
 \end{eqnarray} 
where we used the Gordon identity (\ref{Gordon})
in order to connect the helicity structure of the twist three tensor to the approaches in \cite{BelMuel, Polyakov}.
The first four terms in Eq.~(\ref{tw3_metz}) conserve the proton helicity, while the last four terms 
flip the proton helicity. Furthermore we used the identity 
\[  \overline{U}(p',\Lambda')  \gamma_i  U(p,\Lambda) =  \overline{U}(p',\Lambda')\frac{i \sigma^{ji} \, \Delta_T^j}{M} U(p,\Lambda)  \rightarrow  \left\langle {\bf S}_L \times {\bf \Delta}_T \right\rangle .\]
%
Using the notation of Eq.~(\ref{tw2_operator}), we now consider the matrix elements of the quark twist three operators,
\begin{eqnarray}
\label{tw3_operator}
{\cal O}^q_{\pm \mp}(z) & = & \bar{\psi}\left(-\frac{z}{2}\right) (\gamma_1 \pm i\gamma_2) (1\pm\gamma_5)  \psi\left(\frac{z}{2}\right), 
\end{eqnarray}
which lead to the following expression for the helicity amplitude,
\begin{eqnarray}
\label{helamp_tw3}
A^{tw 3}_{\Lambda' \lambda', \Lambda \lambda} & = \displaystyle\int \frac{d z^- \, d^2{\bf z}_T}{(2 \pi)^3} e^{ixP^+ z^- - i\bar{\bf k}_T\cdot {\bf z}_T} \left. \langle p', \Lambda' \mid {\cal O}_{\lambda' -\lambda}(z) \mid p, \Lambda \rangle \right|_{z^+=0}, 
\end{eqnarray} 
where now, 
\begin{subequations}
\label{tw3_operator2}
\begin{eqnarray}
{\cal O}^q_{\pm \mp}(z) & = & \phi^\dagger_{\pm}\left(-\frac{z}{2}\right) \chi_{\pm}\left(\frac{z}{2}\right)  \pm \, \chi^\dagger_{\pm}\left(-\frac{z}{2}\right) \phi_{\pm}\left(\frac{z}{2}\right).
\end{eqnarray}
\end{subequations}
For the good spinor components, $\phi_\lambda$, helicity and chirality are the same since ($1 \pm \gamma^5$) projects out both $\pm$ helicity and chirality. For the bad components helicity and chirality are opposite as it follows from the fact that $\chi_\lambda$ describes a composite system of a transverse gluon and $\phi_\lambda$. Since the gluon carries helicity but no chirality, by imposing angular momentum conservation one obtains the opposite chirality \cite{KogSop,Jaffe}. 
The net effect of this distinction between helicity and chirality is that the helicity conserving quark correlator at twist three will behave like the chirality odd operator at twist two - the latter flips helicity. That is interpreted as if the genuine twist three correlator has the helicity of the returning quark flipped ($\pm 1/2 \rightarrow \mp 1/2$), while the collinear gluon field is transverse and with compensating helicity ($\pm 1$) \cite{Jaffe}. The implication for the Parity relations, unlike Eq.(\ref{helparity}), is that the reversed helicities are not directly related to the initial set. The twist three correlator does not map directly onto a 2-body process. For this reason there are twice as many vector and axial vector twist three GPDs and TMDs. 

As we have noted, for the non flip quark-proton helicity amplitudes at twist three one finds that the chiral even structures correspond to what would be chiral odd at twist two,
\begin{eqnarray}
A^{ tw 3}_{\Lambda' \pm, \Lambda \pm}  \rightarrow A^{tw 2}_{\Lambda' \pm, \Lambda \mp}.
\end{eqnarray}
By using the operators in Eq.~(\ref{tw3_operator2}), the non-flip helicity amplitudes ($\Lambda = \Lambda'$) can be read off from the hadronic tensors parameterizations as,
\begin{subequations}
\label{A3Lambda}
\begin{eqnarray}
\label{A3Lambda1}
A^{tw 3}_{\Lambda +,\Lambda +} & = &
 \frac{\bar{k}_T^1+i \bar{k}_T^2}{P^+} F_{21} +    \frac{\Delta_1 + i \Delta_2}{P^+} F_{22}  -  \Lambda \frac{\bar{k}_T^1+ i  \bar{k}_T^2}{P^+}   F_{27} - \Lambda \frac{\Delta_T^1+ i  \Delta_T^2}{P^+} F_{28}   \nonumber \\
 & +& \frac{\bar{k}_T^1+i \bar{k}_T^2}{P^+} G_{21}  +   \frac{\Delta_1 + i \Delta_2}{P^+} G_{22} +  \Lambda \frac{\bar{k}_T^1+ i  \bar{k}_T^2}{P^+}   G_{27} +\Lambda \frac{\Delta_T^1+ i  \Delta_T^2}{P^+} G_{28}\\
\label{A3Lambda2}
A^{tw 3}_{\Lambda -,\Lambda -} & = &  \frac{\bar{k}_T^1-i \bar{k}_T^2}{P^+} F_{21} +    \frac{\Delta_1 - i \Delta_2}{P^+} F_{22} +  \Lambda \frac{\bar{k}_T^1- i  \bar{k}_T^2}{P^+}   F_{27} +\Lambda \frac{\Delta_T^1- i  \Delta_T^2}{P^+} F_{28}  \nonumber \\
 & +& \frac{\bar{k}_T^1-i \bar{k}_T^2}{P^+} G_{21} +   \frac{\Delta_1 - i \Delta_2}{P^+} G_{22}  +  \Lambda \frac{\bar{k}_T^1 - i  \bar{k}_T^2}{P^+}   G_{27} +\Lambda \frac{\Delta_T^1- i  \Delta_T^2}{P^+} G_{28}\\
\end{eqnarray}
\end{subequations}
Only two independent combinations of the $A^{tw 3}_{\Lambda \lambda,\Lambda \lambda}$ can be formed, namely, 
\begin{subequations}
\begin{eqnarray}
\label{F22}
\frac{4}{P^+}\left[\frac{\bar{\bf k}_T \cdot   {\bf \Delta}_T}{\Delta_T} F_{21} + \Delta_T F_{22} + \left(\frac{\bar{\bf k}_T \cdot   {\bf \Delta}_T}{\Delta_T} G_{21} + \Delta_T G_{22} \right) \right] & = & A_{++,++}^{tw 3} + A_{+-,+-}^{tw 3} + A_{-+,-+}^{tw 3} + A_{--,--}^{tw 3} \\
 \label{F28}
-\frac{4}{P^+}\left[  \frac{\bar{\bf k}_T \cdot   {\bf \Delta}_T}{\Delta_T} F_{27} +   \Delta_T F_{28} - \left(\frac{\bar{\bf k}_T \cdot   {\bf \Delta}_T}{\Delta_T} G_{27} +   \Delta_T G_{28} \right)\right]  & = & A_{++,++}^{tw 3} +A_{+-,+-}^{tw 3} - A_{-+,-+}^{tw 3} - A_{--,--}^{tw 3}   
\end{eqnarray}
\end{subequations}
where we have taken ${\bf \Delta}_T$ along the $x$-axis without loss of generality. Notice that Eq.(\ref{F22}) corresponds to the unpolarized case yielding the twist two GPD $H$, while Eq.(\ref{F28}) gives the distribution of an unpolarized quark in a longitudinally polarized proton. Owing to the helicity structure of the twist three quark operators discussed above, this combination is now allowed by Parity conservation  \cite{KogSop,Jaffe}.  

Integrating over ${\bf \bar{k}_T}$,  one obtains the twist three GPDs,
\begin{eqnarray}
\label{H2Ttilde}
2\widetilde{H}_{2T} + E_{2T} & = &  \int d^2 {\bf k}_T  \left[ \left(\frac{{\bf k}_T \cdot {\bf \Delta}_T}{\Delta_T^2} \right) F_{21} +  F_{22} \right]
 \\
 \label{E2Ttilde}
\widetilde{E}_{2T} & = & -2  \int d^2 {\bf k}_T \left[ \left(\frac{{\bf k}_T \cdot {\bf \Delta}_T}{\Delta_T^2} \right) F_{27} +  F_{28} \right] \\
\label{H2Ttildep}
2\widetilde{H}_{2T}^\prime + E_{2T}^\prime& = &  \int d^2 {\bf k}_T  \left[ \left(\frac{{\bf k}_T \cdot {\bf \Delta}_T}{\Delta_T^2} \right) G_{21} +  G_{22} \right]
\label{E2Ttildep} \\
\widetilde{E}_{2T}^\prime & = &  -2 \int d^2 {\bf k}_T  \left[ \left(\frac{{\bf k}_T \cdot {\bf \Delta}_T}{\Delta_T^2} \right) G_{27} +  G_{28} \right] 
\end{eqnarray}
in agreement with  Ref.\cite{Metz2}.
In order to proceed, it is important to connect the various notations for the  twist three GPDs which appear classified in the literature in the three main publications, Refs.~\cite{BelMuel,Polyakov,Metz2}, respectively. By using the Gordon relation and  \cite{Kivel} 
\[ i \epsilon^{+i\alpha\beta} \overline{U}(p',\Lambda')  \gamma_\alpha\Delta_\beta \gamma_5 U(p,\Lambda) =  i  \overline{U}(p',\Lambda') \left( \Delta_j \gamma^+ - \Delta_+ \gamma^j \right) \gamma_5 U(p,\Lambda) = 2P^+  \overline{U}(p',\Lambda') \gamma^j U(p,\Lambda),\] 
which follows from the Dirac equation,
we find that all notations, as reported in Table \ref{table:tw3}, are equivalent.
\begin{table}
\begin{tabular}{|c|c|c|c|c|}
\hline
\hline
&&&& \\
Polyakov et al.  \cite{Polyakov} &   $2 G_1$ &   $G_2$  &  $G_3$   &    $G_4$  \\ 
&&&& \\
\hline 
&&&& \\
Meissner et al. \cite{Metz2} & $ 2 \widetilde{H}_{2T} $ & $ \widetilde{E}_{2T}$ & $E_{2T}$ & $ H_{2T}$  \\
&&&& \\
\hline 
&&&& \\
Belitsky et al. \cite{BelMuel}&   $E_+^3$ & $\widetilde{H}_-^3 $   & $H_+^3 + E_+^3$ &  $ \displaystyle\frac{1}{\xi} \widetilde{E}_-^3 $   \\
&&&& \\ 
\hline
\end{tabular} 
\caption{\label{table:tw3}Comparison of notations for different twist 3 GPDs.}.
\end{table}

\vspace{0.3cm}
{\bf 4.} We now turn to the interpretation of the OAM term in the proton's angular momentum sum rule \cite{JM_SR,Ji_SR} which requires, as we show below, 
the twist three helicity amplitudes combination corresponding to $\widetilde{E}_{2T}$, Eq.(\ref{E2Ttilde}). 
While the derivation of the sum rule was carried out along similar lines in both Refs.\cite{JM_SR} and \cite{Ji_SR}, the two approaches essentially differ in that in Ref.\cite{JM_SR} (JM) one has,
\footnote{We do not discuss here the alternative decompositions of angular momentum. For an extensive discussion of this issue see {\it e.g.} Ref.\cite{Waka}. }
\begin{equation}
\frac{1}{2} =  \frac{1}{2} \Delta \Sigma + {\cal L}_q + \Delta G + {\cal L}_g,
\end{equation} 
where ${\cal L}_{q(g)} \rightarrow {\bf r} \times i{\bf \partial}$, {\it i.e.}  corresponds to  canonical OAM, while in Ref.\cite{Ji_SR} (Ji),
\begin{equation}
\label{Ji_SR:eq}
\frac{1}{2} = J_q + J_g =  \frac{1}{2} \Delta \Sigma + L_q + J_g,
\end{equation}
where $L_q \rightarrow  {\bf r} \times i{\bf D}$ includes dynamics through the covariant derivative. $J_g$, the gluons total angular momentum contribution to Eq.(\ref{Ji_SR:eq}) was originally not split into its separate intrinsic and orbital components, in order to satisfy gauge invariance.  We note, however, that in Ref.\cite{Waka} a separation into all four parts (the
orbital angular momenta and intrinsic spins of quarks and gluons) was proposed that is gauge invariant
for the longitudinal nucleon spin. This question, and similarly  the feasibility  of a gauge invariant separation into all four components for the transverse nucleon spin, are still a matter of debate and beyond the scope of this paper.

In Ref.\cite{Ji_SR} the quarks and gluons angular momentum components were identified with observables obtained from Deeply Virtual Compton Scattering (DVCS) type experiments. Both $J_{q(g)}$ and $L_q$ can therefore be measured owing to the well known relation involving twist two GPDs, 
\begin{eqnarray}
 \frac{1}{2} \int_{-1}^1 dx \, x(H_{q(g)}(x,0,0) &+&E_{q(g)}(x,0,0)) = J_{q_(g)} \rightarrow \nonumber \\
L_q =  \frac{1}{2} \int_{-1}^1 dx \, &x& (H_{q}(x,0,0)+E_{q}(x,0,0)) - \frac{1}{2} \int_{-1}^1 dx \, \widetilde{H}(x,0,0)
\end{eqnarray}
(the arguments of the GPDs are $(x,\xi=0,t=0)$.
What is crucial here is that in a subsequent development Polyakov {\it et al.} \cite{Polyakov} derived a sum rule for the twist three vector components,
\begin{eqnarray}
\label{polyakov:eq}
\int dx \, x \, G_2^q(x,0,0)  =  \frac{1}{2} \left[ - \int dx  x  (H^q(x,0,0) + E^q(x,0,0)) + \int dx \tilde{H}^q(x,0,0) \right] 
\end{eqnarray}
from which it appears that the second moment of $G_2 \equiv \widetilde{E}_{2T}$ (see Table \ref{table:tw3}) represents the quarks' OAM.

By unraveling the helicity structure of $\widetilde{E}_{2T}$ (or equivalently, $G_2$) in Eqs.(\ref{F28}) and (\ref{E2Ttilde}) we were able to show that OAM is measured by a twist three contribution which corresponds to the Lorentz structure $\sigma^{ij} \Delta_j$ \cite{Metz1}, or, in terms of 3-vectors,  to a transverse direction (${\bf S}_L \times {\bf \Delta}_T$). This is at variance with the distribution of an unpolarized quark in a  longitudinally polarized proton appearing at twist two in \cite{LorPas,Yuan}.
%

Our finding is in line with the recent observation that the same twist three contribution is fundamental for solving the issue of defining the quarks and gluons angular momentum  decomposition within QCD \cite{Ji_Xiong,Hatta}.   
An important question remains to be explained of whether $G_2$ is related to the distribution of canonical angular momentum, ${\cal L}_{q}$.  
In order to address this question we notice that,
\begin{eqnarray}
L_q(x) & = & L_q^{WW}(x) + \overline{L}_q(x) 
\label{Ldyn}
\\
{\cal L}_{q}(x) & = & L_q^{WW}(x)  + \overline{\cal L}_{q}(x),
\label{Lcan}
\end{eqnarray} 
where $L_q^{WW}(x)$  is the  Wandzura Wilczek (WW) contribution, $\overline{L}_q(x)$ and $\overline{\cal L}_{q}(x)$ are the genuine twist three terms. Eqs.(\ref{Ldyn},\ref{Lcan})  are consistent with the observation  that ${L}_q(x)$ and ${\cal L}_{q}(x)$ admit the same WW part, while they differ in their genuine twist three contribution \cite{Hatta}. In other words, 
while in the WW limit the two OAM distributions coincide, their differences depend on final state interactions contained in this case in the genuine twist three terms (notice, in particular, that $\int dx   \overline{L}_q(x) = 0$). 


Only the WW contribution to $G_2$ and $L_q$, obtained by taking the forward limit of the twist three GPDs \cite{BelMuel,Kivel,Hatta},
contributes to the sum rule in Eq.(\ref{polyakov:eq}). One has,
\begin{equation}
L_q^{WW}(x,0,0) =  x \int_x^1 \frac{dy}{y} (H_q(y,0,0)+E_q(y,0,0)) - x \int_x^1 \frac{dy}{y^2} \widetilde{H}_q(y,0,0),
\label{eq:WW}
\end{equation}    
An alternative argument was given in \cite{Ji_Xiong,Hatta}, and demonstrated with a specific physical example in \cite{Burkardt_torque}, where, by treating OAM in a similar way to Single Spin Asymmetries (SSA's), the difference between $L_q$ and ${\cal L}_{q}$ has been attributed, within the TMD factorization picture,  to final state interactions via the specific behavior of the gauge links in the two cases. We remind that this approach relies on Wigner distributions, whereas the issue of its connection to OPE needs further discussion.

\begin{figure}
\includegraphics[width=9cm]{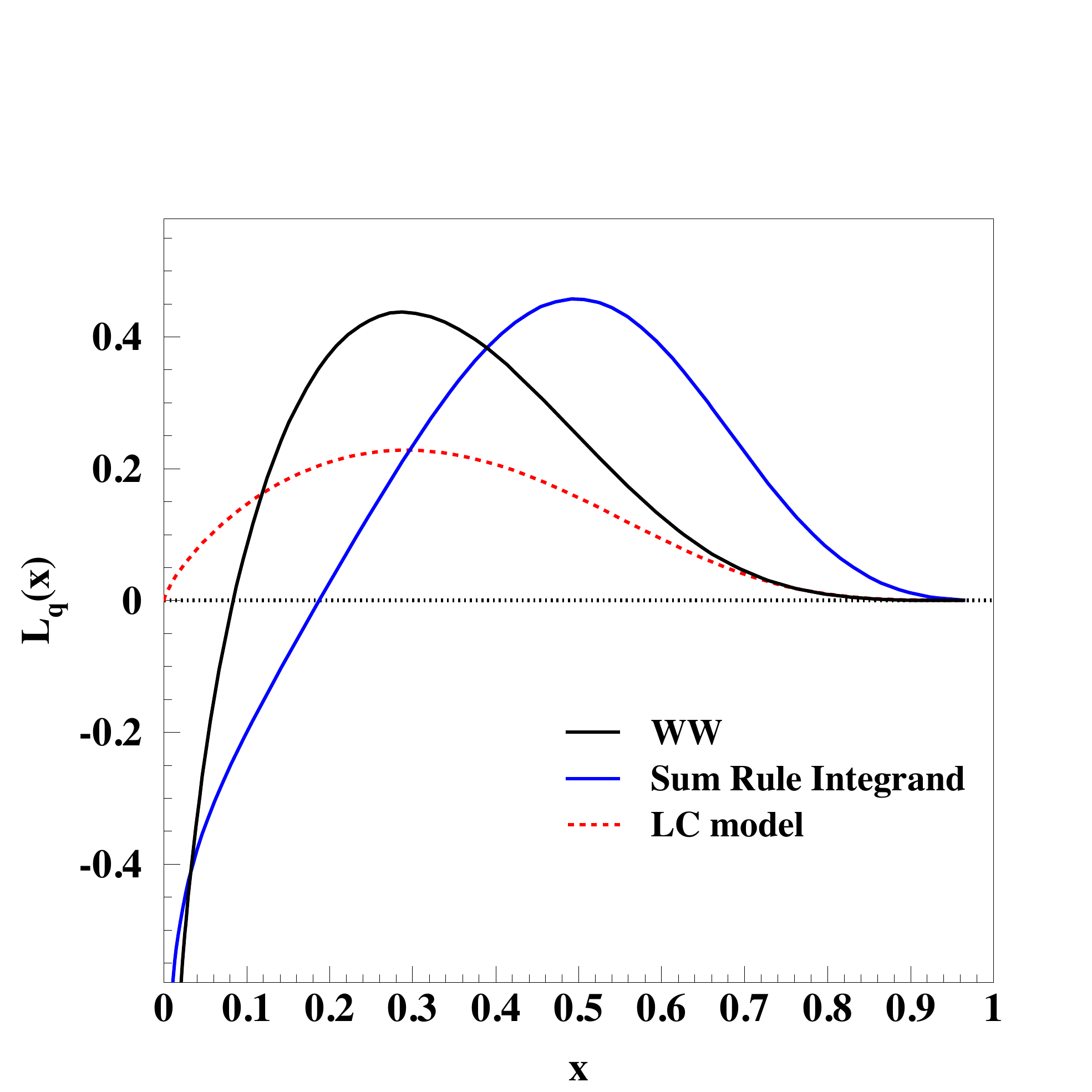}
\caption{(Color online) Unintegrated OAM for the $u$ quark calculated in the Wandzura Wilczek approximation Eq.(\ref{eq:WW}), compared to the integrand in Ji's sum rule, Eq.(\ref{Ji_SR:eq}), evaluated using the reggeized diquark model of Refs.\cite{GGL,GGL_FF} . Results are compared to ${\cal L}_u$ obtained using the same input wave functions from Refs.\cite{GGL,GGL_FF} in the partonic picture given in  Refs.\cite{BurkardtBC} (labeled LC in the graph). }
\label{fig:oam}
\end{figure}

In order to illustrate the feasibility of OAM measurements, in Fig.\ref{fig:oam} we show the unintegrated OAM for the $u$ quarks,  within the WW approximation, Eq.(\ref{eq:WW}). 
We show for comparison the integrand in Ji's sum rule, Eq.(\ref{Ji_SR:eq}), and an evaluation obtained in a partonic picture following Refs.\cite{BurkardtBC}. All curves were obtained using the same parametrization for the GPDs $H, E, \widetilde{H}$ of Ref.\cite{GGL,GGL_FF}. The WW and Ji-integrand curves integrate to the same value of $L_q = 0.13$. The curve from Ref.\cite{BurkardtBC} integrates to ${\cal L}_q = 0.11$. 

Next we propose an observable that gives a direct experimental access to the quark OAM.  
The GPDs contributions to the amplitudes for Deeply Virtual Compton Scattering (DVCS) experiments were computed up to twist three in Refs.\cite{BelMuel,Kivel} (see also \cite{BelRad}). 
By referring to  the notation of  Ref.\cite{BelMuel} for the CFFs we see that the contribution of $G_2$ is  found in the 
singularity free combination given by 
\begin{equation}
\widetilde{\mathcal{H}}^{eff} = -2 \xi \left( \frac{1}{1+\xi} \widetilde{\mathcal{H}} + \widetilde{\mathcal{H}}^+_3 - \widetilde{\mathcal{H}}^-_3 \right),
\end{equation}
where (Table \ref{table:tw3}),
\begin{eqnarray}
\widetilde{\mathcal{H}} & = & C^+ \otimes \widetilde{H},  
\;\;\;\;
\widetilde{\mathcal{H}}_3^+   =  C^+\otimes  \widetilde{E}_{2T}^\prime, 
\;\;\;\;
\widetilde{\mathcal{H}}_3^-  =  C^-\otimes  \widetilde{E}_{2T}, 
\end{eqnarray}
and,
\begin{equation}
C^\pm = -\frac{1}{x-\xi + i \epsilon} \pm  \frac{1}{x+\xi - i \epsilon}
\end{equation}
In order to extract $G_2 \equiv \widetilde{E}_{2T}$ from experiment one needs to first of all single out the observables sensitive to $\widetilde{\cal H}^{eff}$.  These are the azimuthal
asymmetries for DVCS on a longitudinally polarized proton, namely 
the single spin asymmetry averaged over all
beam polarizations,  $A_{UL}$,  and the double spin asymmetry where
both beam and target are polarized, $A_{LL}$. Both $A_{UL}$, and $A_{LL}$ have been recently measured  \cite{HERMES, pisano_privcomm}. 

In this Letter we show our calculation of the twist two and twist three contributions to $A_{UL}$  \cite{BelMuel},
\begin{equation}
A_{UL} = \frac{N_{s_z=+}-N_{s_z=-}}{N_{s_z=+}+N_{s_z=-}} 
\end{equation}
where $N_{s_z=\pm}$ is a measure of the number of scatterings on a proton with longitudinal spin, $s_z= \pm 1/2$.
The dependence of $A_{UL}$ on the the angle $\phi$, or the azimuthal angle between the
lepton plane and the plane of the virtual and real photons can be written keeping terms up to twist three as,
\begin{equation}
A_{UL} = \frac{a \sin \phi + b \sin 2 \phi}{c_0+ c_1 \cos \phi + c_2 \cos 2 \phi} 
\end{equation}
where the coefficients for the total unpolarized cross section in the denominator,  $c_0$, $c_1$, and $c_2$ are given by combinations of the Bethe-Heitler (BH), DVCS, and BH-DVCS interference terms \cite{BelMuel}. 
The coefficients in the numerator, also displayed in \cite{BelMuel}  contain the GPDs of interest in our study namely, 
\[ a \approx s^{\mathcal{I}}_{1,LP} \propto F_1(t) \Im m  \widetilde{\mathcal{H}} \] 
and 
\[b \approx s^{\mathcal{I}}_{2,LP} \propto F_1(t)  \Im m \widetilde{\mathcal{H}}^{eff}, \]
where $F_1(t)$ is  the Dirac form factor.

\begin{figure}
\includegraphics[width=9cm]{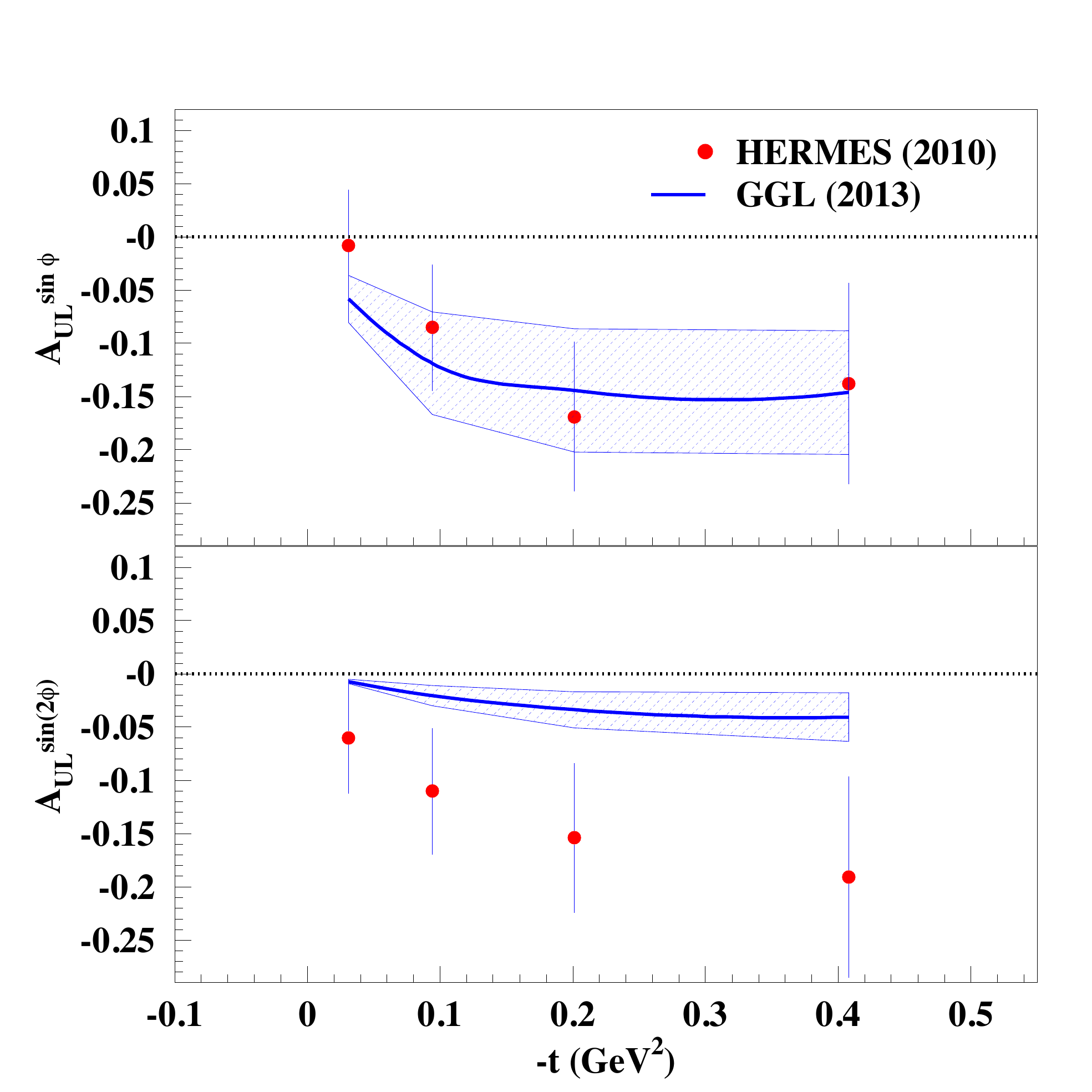}
\caption{(Color online) The asymmetry $A_{UL}$  twist two ($\sin \phi$) and twist three ($\sin 2 \phi$) modulations plotted vs. the momentum transfer squared $-t$, compared to HERMES data \cite{HERMES} at the Bjorken $x$ and scale $Q^2$ of the data. 
The blue bands represent the predictions from  the GPD model of \cite{GGL,GGL_FF} denoted by GGL in the legend,  including the error from the model's parameters variations calculated  in WW approximation. }
\label{fig:oam2}
\end{figure}

 
In Fig.\ref{fig:oam2} we show the asymmetry values plotted vs. the momentum transfer squared $-t$, compared to HERMES data \cite{HERMES} at the Bjorken $x$ and scale $Q^2$ of the data. Both the twist two ($\sin \phi$) and twist three ($\sin 2 \phi$) modulations are shown. 
The blue bands represent the predictions from  the GPD model of \cite{GGL,GGL_FF} including the error from the model's parameters variations calculated  in WW approximation. 
As we can see from the figure the $\sin 2\phi$ modulation, dominated by the $\tilde{\cal H}$ Compton form factor, is sizable: an extraction of $G_2$ is then possible.
More accurate data analyses that will allow us to better single out this term will be available soon \cite{pisano_privcomm}. 

Finally, we notice that, as shown recently in \cite{Hatta},  both the canonical \cite{JM_SR} and Ji's OAM  admit the same WW approximated form, while their genuine twist three contributions differ. 

\vspace{0.3cm}
\noindent {\bf 5.}  
In conclusion, we have proposed the first experimental access to the quark OAM, through twist three GPDs. With the corresponding data from HERMES and the soon available data from JLab, $L_q$ could be extracted. 

Our suggestion for a direct OAM measurement originates from an interpretation of the helicity structure of GTMDs and GPDs that identifies the relevant spin projections for this 
quantity. In particular we show that OAM is determined by a transverse spin correlation at twist three.
 
The non-zero $F_{14}, G_{11}$ cannot directly be related to the single longitudinal polarizations of either the quarks or the nucleons within the transverse momentum distributions, and final state interactions should be introduced. Efforts to extract dynamical information from model calculations of these will lead to deceptive results.
Because the quarks are off their mass shell before imposing any particular model, the counting of independent helicity amplitudes is complicated leading to a doubling of the number of helicity states \cite{FGM}. Starting from this observation we proposed a QCD approach where: 1) single longitudinal polarizations observables can be derived; 2) they involve twist three distributions. Our approach is complementary to the one in Ref.\cite{Burkardt_torque} that was derived using TMD factorization.

Our most important result is perhaps in dispelling the notion that what is believed to be the orbital angular momentum component of the nucleon spin sum rule cannot be observed directly in hard scattering experiments. Both the JM and Ji decompositions correspond to twist three contributions, and their validity can be tested by measuring twist three GPDs. These observables can be obtained from both HERMES data \cite{HERMES} and in forthcoming Jefferson Lab analyses \cite{pisano_privcomm}.   

\vspace{0.3cm}
We are grateful to  Silvia Pisano for fruitful discussions as well as information about the DVCS observables. This work was funded in part by the Belgian Fund F.R.S.-FNRS via the contract of Charge de recherches (A.C.), and by U.S. D.O.E. grant DE-FG02-01ER4120 (S.L., A.R.). A.R. thanks the INFN Laboratori Nazionali di Frascati for support.

\end{document}